# Self-waveguiding of relativistic laser pulses in neutral gas channels

L. Feder*, B. Miao*, J.E. Shrock, A. Goffin, and H.M. Milchberg[§]

*Institute for Research in Electronics and Applied Physics, University of Maryland, College Park, MD 20742*
[§]*milch@umd.edu*

**Abstract:** We demonstrate that an ultrashort high intensity laser pulse can propagate for hundreds of Rayleigh ranges in a prepared neutral hydrogen channel by generating its own plasma waveguide as it propagates; the front of the pulse generates a waveguide that confines the rest of the pulse. A wide range of suitable initial index structures will support this "self-waveguiding" process; the necessary feature is that the gas density on axis is a minimum. Here, we demonstrate self-waveguiding of pulses of at least $1.5 \times 10^{17}$ W/cm$^2$ (normalized vector potential $a_0 \sim 0.3$) over 10 cm, or ~100 Rayleigh ranges, limited only by our laser energy and length of our gas jet. We predict and observe characteristic oscillations corresponding to mode-beating during self-waveguiding. The self-waveguiding pulse leaves in its wake a fully ionized low density plasma waveguide which can guide another pulse injected immediately following; we demonstrate optical guiding of such a follow-on probe pulse.

## 1. INTRODUCTION

Laser wakefield acceleration (LWFA) has been the subject of intensive worldwide research for its promise of producing high quality electron bunches in the range of hundreds of GeV to TeV for high-energy physics [1–3] A single acceleration stage of such a device will require a guiding structure to maintain an intense drive laser pulse at relativistic intensities over many Rayleigh ranges of propagation with high group velocity and low electromagnetic leakage loss [4]. Practically, this requires meter-scale low density plasma waveguides.

The first plasma waveguides were formed by the hydrodynamic expansion of a high temperature plasma formed by inverse bremsstrahlung (IB) heating of a gas driven by a ~100 ps axicon-generated Bessel beam pulse, where the cylindrically expanding shock wave formed the waveguide "cladding" with peak density $N_{e,max}$ surrounding a lower density "core" of on-axis density $N_{eo}$ [5]. Later experiments with lower density gases used heating by ultrashort pulse induced optical field ionization (OFI) to drive hydrodynamic shock expansion [6–8] However, the initial maximum temperature $T_{e0}$ of OFI plasmas is limited to the ponderomotive potential at the time of ionization, which is ~10 eV for hydrogen at its ~$10^{14}$ W/cm$^2$ ionization threshold [9] This limits the waveguide depth ($N_{e,max} - N_{eo}$) and cladding thickness, giving electromagnetic leakage distances $L_{1/e} < \sim 1$ cm [10], far shorter than needed for > GeV acceleration.

Our recent experiments [11] have demonstrated a solution to this weak confinement: a second, higher order Bessel beam imprints the cladding by ionizing the neutral gas just outside the initial plasma column, creating low density, low loss plasma waveguides. Here, we investigate another, quite different approach: "self-waveguiding" of relativistic laser pulses in prepared neutral gas refractive index profiles. The physics of self-waveguiding is quite distinct from the well-known process of relativistic self-guiding, where relativistic self-focusing leads to electron cavitation by the laser ponderomotive force [1,2]. In self-waveguiding, the front of the pulse generates a real plasma waveguide that confines the rest of the pulse—the front of the pulse does the optical field ionization 'work' of generating the waveguide, and then the rest of the pulse is channeled by it. Relativistically intense pulses are not required, but pulses must exceed the optical field ionization threshold of the target gas. The process requires an initially prepared elongated refractive index structure whose necessary feature is a gas density minimum on axis. This requires impulsive heating of the gas along an extended path; we show that plasma generation by optical field ionization of hydrogen gas by a Bessel beam pulse supplies sufficient heating. However, an initial plasma is necessary only to the extent that it heats the gas to generate the axial neutral density depression. We note that prior work by Morozov *et al*. [12] had shown that the leading edge of a pulse injected into an incompletely ionized plasma waveguide can ionize the peripheral neutrals and lead to guiding over a few mm.

---

* these authors contributed equally to this work

Here, we present detailed experiments and simulations exploring self-waveguiding of weakly relativistic pulses over 10 cm (100 Rayleigh ranges), and measuring in detail the guided mode structure and time resolved refractive index profiles of plasma and neutral hydrogen. We show that during self-waveguiding, guided mode beating can be significant. A self-waveguided pulse can be also be viewed as a vehicle for generating the plasma waveguide that is left in its wake, suitable for guiding an even more intense follow-on pulse to drive applications such as LWFA. To demonstrate this principle, we demonstrate guiding of a probe pulse that closely follows the self-waveguided pulse.

## 2. EXPERIMENTAL SETUP

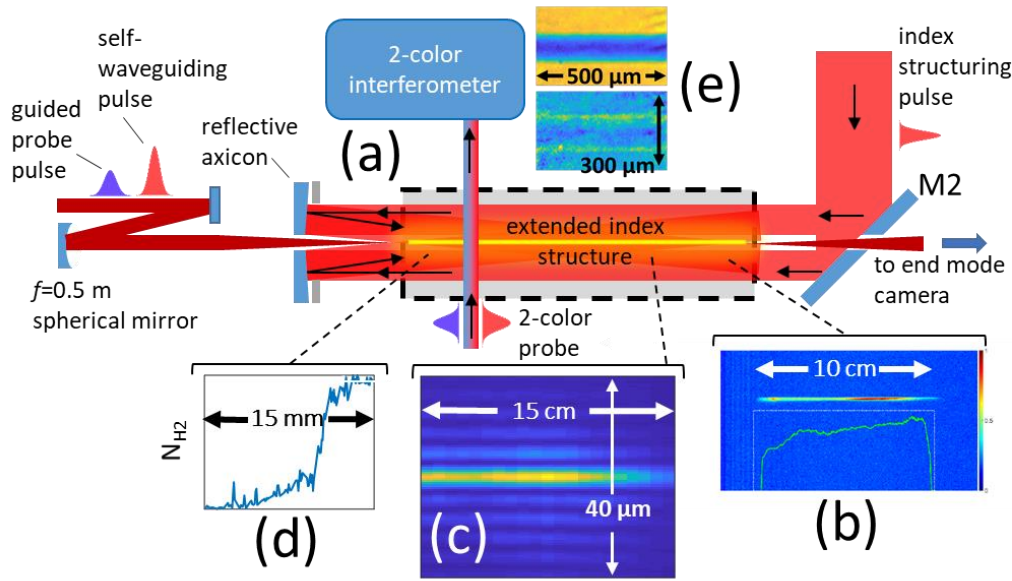

**Figure 1. (a)** Experimental setup, showing index structuring Bessel beam pulse ($J_0$ pulse) (100 mJ, 75 fs linearly polarized, $\lambda$=800 nm) focused by a reflective axicon with base angle $\alpha = 3°$, over a 10 cm long supersonic hydrogen gas jet. After a delay $\Delta t_{inj}$, a 40 fs, 10-90 mJ, linearly polarized pulse (self waveguiding pulse) is focused through a hole in the axicon at $f/40$ and injected into the entrance of the index structure. The exit of the index structure is imaged by an end mode CCD camera. A follow-on pulse (guided probe pulse) is guided by the plasma waveguide generated by the self-waveguiding pulse. **(b)** H$_2$ gas density profile and fluorescence from plasma recombination after optical field ionization by the $J_0$ pulse. **(c)** Magnified axial scan of the $J_0$ beam profile. **(d)** ~2mm density falloff at the entrance of the jet. **(e)** 2-colour ($\lambda$=400 nm and 800 nm) interferometric probe pulse is directed transversely through the index structure and into a folded wavefront interferometer, with extracted phase maps from the electron density (top) and neutral (bottom) profiles shown [14].
2

Figure 1 shows the experimental setup. A 75 fs, 100 mJ, 10 Hz Ti:Sapphire pulse (index structuring pulse) is focused by a reflective axicon with a $d_h = 1.27\ cm$ diameter central hole and base angle $\alpha = 3°$ (Fig. 1(a)) over a 10-cm long pulsed supersonic H$_2$ gas jet (Fig. 1(b)), generating a refractive index structure composed of plasma and neutral hydrogen. The length of the resulting Bessel beam ($J_0$ pulse) focus is $L_B \approx (D - d_h)/2\tan(2\alpha) \approx 16$ cm, where $D$ =4.5 cm is the beam diameter, overfilling the jet 0.8 mm above its 10 cm long slot orifice, where the local hydrogen molecular density was adjusted in the range $N_{H2} = 10^{18} - 10^{19}$ cm$^{-3}$.

A central 2D slice of the axially scanned $J_0$ pulse focal image is shown in Fig. 1(c). After a delay $\Delta t_{inj}$, a 40 fs, 10-90 mJ, linearly polarized pulse (self waveguiding pulse) is focused through a hole in the axicon at $f/40$ and injected into the entrance of the refractive index structure. The exit of the index structure is imaged by an end mode CCD camera. For experiments in a pump-probe configuration, a slightly delayed probe pulse is injected into the guide generated by the self-waveguiding pulse. The gas sheet is generated by a 10 cm long supersonic slit nozzle (Mach ~4), with hydrogen gas fed into the nozzle by fast pulsed valves. The valve open time was 3 ms to ensure equilibrium gas flow. The gas density fall-off scale length at the ends of the gas sheet is ~2 mm (Fig. 1(d)) A 2-colour interferometric probe pulse (λ=400 nm and 800 nm) passes transversely through the index structure and into a folded wavefront interferometer, enabling separate extraction [13] of the neutral hydrogen and electron density contributions to the refractive index profile, with sample phase maps shown in Fig. 1(e). Knowledge of the individual contributions to the composite index profile is crucial for engineering the self-waveguiding process.

## 3. RESULTS AND DISCUSSION

To begin, we assess the ability of an OFI-heated hydrogen plasma to confine and guide a low intensity end-injected laser pulse by examining the plasma's transverse refractive index profile. The 2-colour interferometric probe pulse was passed transversely through the $J_0$ pulse-induced index structure 1 cm past its entrance (the end closest to the axicon). Figure 2(a) shows time resolved electron density and neutral hydrogen density profiles from Abel-inversion of extracted interferometric phase images, where the error bars, here and in subsequent plots, are from phase error propagation in Abel inversion [14]. Just after arrival of the $J_0$ Bessel beam pulse at $t = 0$, the electron density profile is peaked on axis at ~$3.5 \times 10^{18}$ cm$^{-3}$, after which the peak drops to ~$5 \times 10^{17}$ cm$^{-3}$ over 4 ns as the profile radially expands. For our 100 mJ $J_0$ pulse and 50 torr hydrogen target, the electron density profile never develops an on-axis minimum and shock wall cladding characteristic of a hydrodynamic plasma waveguide [3–8], and thus it would be unable to optically guide a laser pulse. The neutral hydrogen density profile at $t = 0$ (obtained from $N_0 - N_e(r, t = 0)$) is depleted on axis and at later times develops a steepened annular enhancement as peripheral neutral hydrogen is "snowplowed" and shock-compressed by the cylindrically expanding electron density column. The peak neutral density in this annular shock wall is $> \sim 2N_0$ (and $> \sim 3N_0$ for 100 torr hydrogen) and $< \sim 20N_{e0}$. The electron density column expands and cools as a cylindrical blast wave expansion $\langle r_{plasma} \rangle = \xi_0 (\varepsilon_0/\rho)^{1/4} t^{1/2}$, where $\langle r_{plasma} \rangle$ is the density-weighted average plasma radius, $\varepsilon_0$ ($\propto T_{e0}$) is the $J_0$ pulse energy deposition per unit length, $\rho$ is the initial mass density of the hydrogen gas, and $\xi_0$ is a dimensionless prefactor of order unity. Figure 2(b) plots the time evolution of $\langle r_{plasma} \rangle$ and the peak of the neutral annulus, with a fit to $\langle r_{plasma} \rangle \propto t^n$ giving $n = 0.5 \pm 0.4$. The plasma temperature, also plotted in Fig. 2(b), is calculated from $(d\langle r_{plasma} \rangle/dt) \approx (\gamma_c T_e/m_i)^{1/2}$, where $\gamma_c$ is the specific heat ratio and $m_i$ is the proton mass. After a few nanoseconds, the plasma expansion stagnates as it cools to $> \sim 1$ eV, with its central density reduced by $\sim 10\times$ and the column expanded by $\sim 5\times$. The neutral annulus speed is $v_N = 7.7\times 10^5$ cm/s (Mach ~6 in hydrogen [15])), giving a neutral hydrogen shock temperature $T_H \sim 0.4$ eV, consistent with the approach to electron-ion local temperature equilibration. The neutral annulus is clearly a neutral hydrogen shock wave that separates from the stagnating plasma and propagates outward at its local sound speed. We simulate the generation of a neutral hydrogen shock initiated by a central 10 eV hydrogen plasma, using a 1D Lagrangian hydrocode described in [16], with shock position vs time overlaid on Fig. 2(b). The computed shock velocity, $\sim 7\times 10^5$ cm/s, is consistent with the measurements.

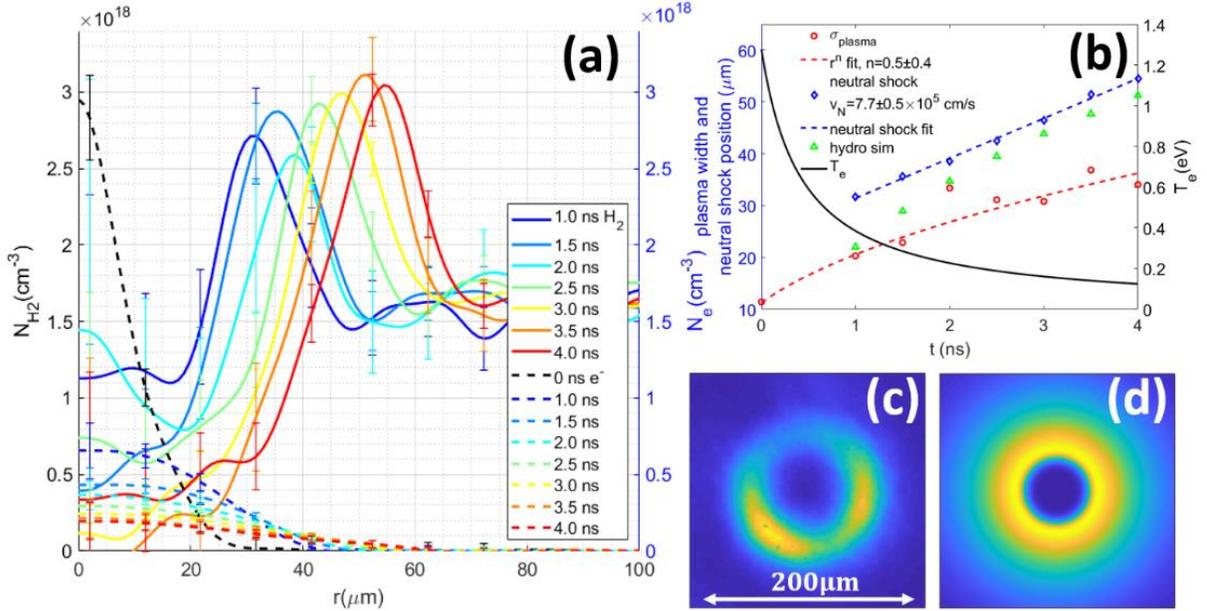

**Figure 2.** (a) Plasma and neutral gas density profile evolution measured by 2-color transverse interferometry. (b) $J_0$ pulse-induced RMS plasma column radius $\sigma_{plasma}$ and neutral shock position vs time. The plasma temperature is computed from the plasma expansion (see text). The neutral shock speed $v_N \sim 7.7\times 10^5$ cm/s, calculated from the plotted points, indicates a gas temperature ~0.4 eV. Green triangles: neutral shock position vs. time from hydrodynamic simulation. (c) Ring mode image of λ=400 nm probe pulse trapped and guided in neutral shock cylindrical annulus. Hydrogen backfill, 50 torr. (d) Beam propagation method (BPM) simulation [17] using conditions of (c). Error bars, here and in subsequent plots, are from phase error propagation in Abel inversion [14].

Another way to image the annular neutral shock wall is to guide an axial probe pulse in it; we have discussed this previously in the context of cylindrical single-cycle acoustic fronts shed by femtosecond filaments in air [18]. Here, at $\Delta t_{inj} = 2.5$ ns delay after the $J_0$ pulse, we injected a λ=400 nm, <100 µJ probe pulse at the entrance of the 10 cm long structure and imaged the guided mode at the exit, with the result being the ring mode shown in Fig. 2(c). This is in good agreement with beam propagation method [17] simulations, shown in Fig. 2(d), for λ=400 nm Gaussian beam injection into the measured 2.5 ns delay profile of Fig. 2(a): the ring mode is laser light trapped by the enhanced density of the cylindrical shock wave.

At first glance, the refractive index structure described so far—a long, low density plasma cylinder surrounded by a cylindrical annulus of enhanced neutral density—appears unsuitable as a guiding structure for high intensity laser pulses. However, the ability of the cylindrical shock wave to guide a low intensity pulse (Fig. 2(c)) ensures that the transverse wings of the leading edge of a high intensity pulse will be guided. Later time slices of the pulse envelope might then ionize the neutral annulus, generating–on the fly–a suitable plasma waveguide cladding for the remainder of the pulse envelope.

Results of an experiment to test this scenario are shown in Fig. 3. Here, a high-power ($P < 5$ TW, linearly polarized, $\lambda = 800$ nm, $\tau_{FWHM} = 40$ fs) laser pulse was focused by a 0.5 m focal length spherical mirror (at $f/40$) to a $w_0 = 25$ μm spot (Rayleigh range $z_0 = \pi w_0^2/\lambda \sim 2.5$ mm) at the entrance of the index structure generated by the $J_0$ pulse. Here, the uncorrected focal spot had ~30% energy within a radius $w_0$. The maximum peak intensity on axis is $I_0 = 2 \times 10^{17}$ W/cm$^2$ and falls to ~$10^{14}$ W/cm$^2$ at a radial distance $r \sim 50$ μm, sufficient for the far transverse wings to ionize any of the annular neutral gas profiles plotted in Fig. 2(a). The electron density profiles in Figure 3(a) show the effect of increasing the energy of a pulse injected at a delay $\Delta t_{inj} = 2.5$ ns (with respect to the $J_0$ pulse) into the refractive index structure shown in Fig. 2(a). Here, the interferometric probe pulse sampled the plasma 2 cm downstream (~$8z_0$) from the structure entrance. It is seen that as pulse energy increases, a plasma waveguide structure forms and converges to a near-final form, consistent with the initial neutral density annulus, except for slight increases in cladding height and width. For fixed injected pulse energy (88 mJ) and $\Delta t_{inject} = 2.5$ ns, Fig. 3(b) shows the increasing plasma density of the cladding as gas pressure is increased. After intense pulse injection, the annular cylinder of new plasma is driven inward and outward by the pressure gradients on the inside and outside of the annulus, distorting the guiding structure, as seen in Fig. 3(c). The inward pressure wave even causes plasma enhancement (compression) on axis, as seen in prior work [19]. The effect of end-injection at various delays after the $J_0$ pulse is shown in Fig. 3(d)— at increasing $\Delta t_{inj}$, the annular zone of additional ionization moves radially outward, following the location of the neutral shock.

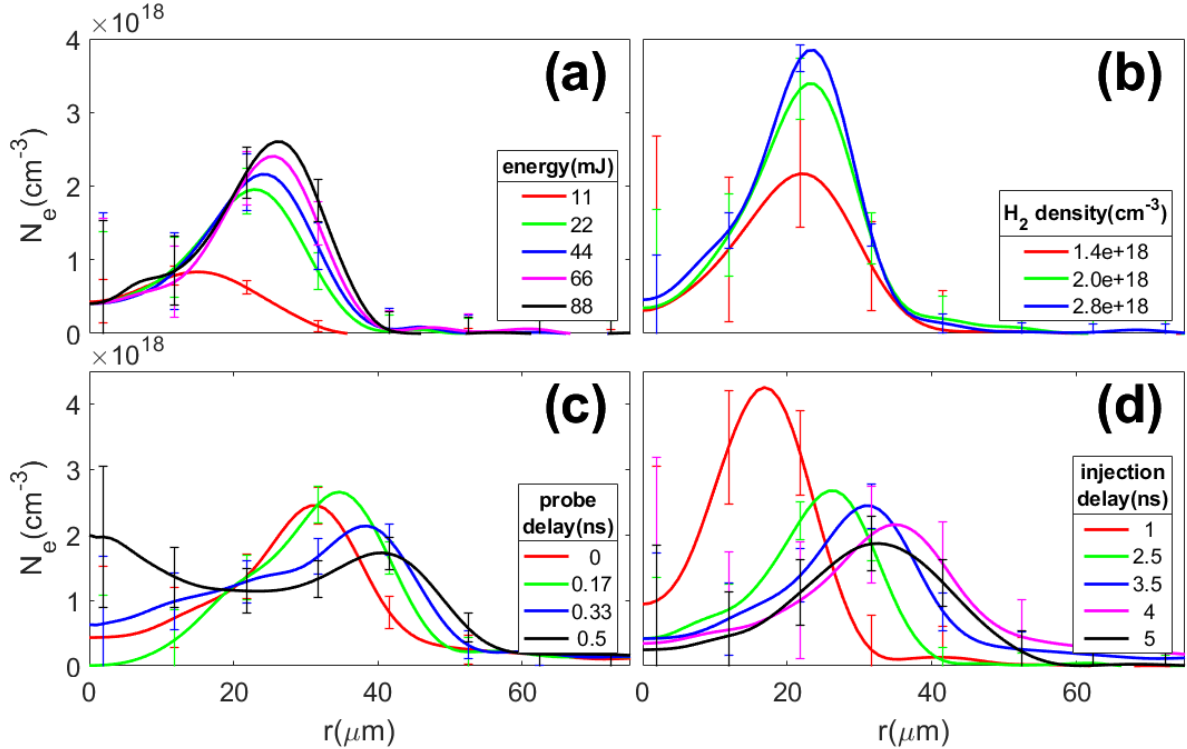

**Figure 3. (a)** Electron density profiles generated at $z = 2$ cm vs injected pulse energy. $\Delta t_{inj} = 2.5$ ns, $N_{H2} = 1.6 \times 10^{18}$ cm$^{-3}$. **(b)** $\Delta t_{inj} = 2.5$ ns, injection beam energy = 88 mJ. **(c)** $\Delta t_{inject} = 3.5$ ns, $N_{H2} = 1.6 \times 10^{18}$ cm$^{-3}$, injection beam energy = 88 mJ. **(d)** $N_{H2} = 1.6 \times 10^{18}$ cm$^{-3}$, injection beam energy = 88 mJ. All panels: Probe delay (with respect to injection pulse) = 5 ps.

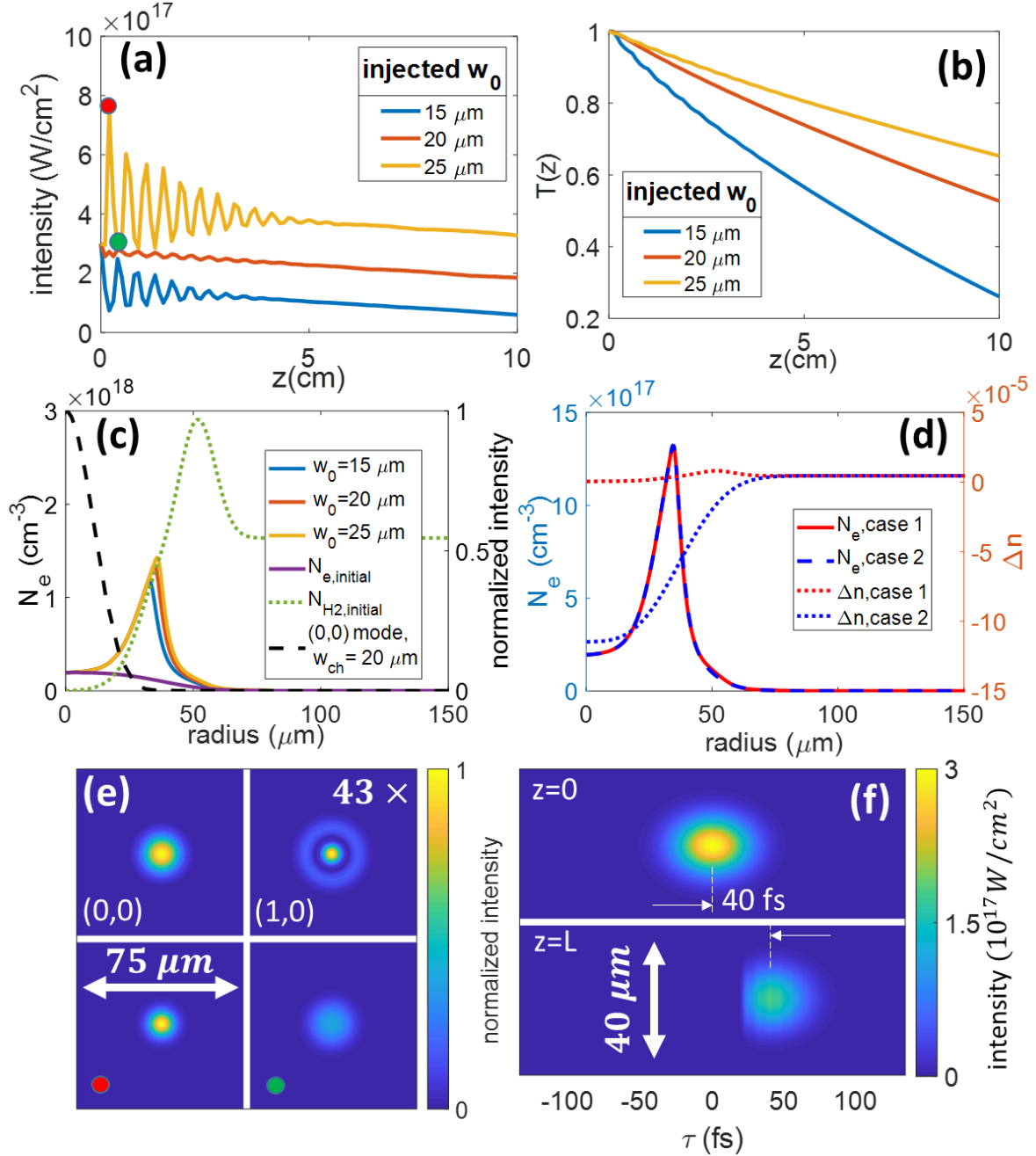

**Figure 4.** Peak input intensity at $z = 0$ is $I_0 = 3 \times 10^{17}$ W/cm². **(a)** Peak laser intensity $I(r = 0, z, \tau_{peak})$ vs. $z$ for various injection pulse waists $w_0$. **(b)** z-dependent transmission efficiency, $T(z) = \varepsilon_i^{-1} \int d\tau d^2 \mathbf{r}_\perp I(\mathbf{r}_\perp, z, \tau)$, for the 3 input modes of (a), with variables defined in the text. **(c)** Similar $N_e(r)$ profiles generated by varying $w_0$ mode injection into the initial index profile shown (characterized by $N_{e,\text{initial}}$ and $N_{H2,\text{initial}}$). **(d)** Demonstration that a range of possible initial refractive index structures can support self-waveguiding, with almost identical plasma waveguides generated. The main feature necessary for self-waveguiding is an on-axis neutral gas density depression. **(e)** Top 2 panels: (0,0) and (1,0) quasi-bound intensity modes of plasma waveguide in (b). The (1,0) mode is shown scaled in intensity by 43 ×. Bottom 2 panels: propagating mode at $z$ locations of intensity maximum and minimum (green and red dots in (a)). **(f)** Pulse intensity envelopes in $(r, \tau)$ at the entrance ($z = 0$, top panel) and exit ($z = L = 10$ cm, bottom panel) of the index structure. The plots show pulse head erosion and group velocity delay in the self-generated waveguide. The pulse propagates right to left.

While Fig. 3 clearly shows that the injected pulse transforms the initial refractive index profile into a suitable plasma waveguide structure at an axial location $8z_0$ past the entrance, the pulse has evidently undergone guided propagation to that point and beyond. To obtain physical insight into the propagation dynamics, we performed simulations of end-injection of intense pulses into a range of refractive index profiles. We use a GPU-based implementation of the unidirectional pulse propagation equation (UPPE) [14,18,20] applied to an intense λ=800 nm, $\tau = 40$ fs FWHM Gaussian pulse of adjustable waist $w_0$ ($1/e^2$ intensity radius) placed at the structure entrance at $z = 0$.

Figures 4 and 5 show simulations for peak intensities $I_0 = 3 \times 10^{17}$ W/cm$^2$ and $1.5 \times 10^{17}$ W/cm$^2$, respectively, injected at the structure entrance at $z = 0$. These intensities bracket our estimated peak injected intensity of $2.2 \times 10^{17}$ W/cm$^2$ and, as we will see, illustrate the intensity sensitivity of the self-waveguiding process. We note that simulations of even higher intensities would require particle-in-cell simulations: as $a_0 \to 1$, the laser ponderomotive force begins to modify the electron density profiles. Our goal for this paper's simulations is to study the nonlinear optics of self-waveguiding and model our experiments for $a_0 < \sim 0.3$, for which an UPPE-based model is quite adequate.

Figure 4(a) shows the peak pulse intensity $I(r = 0, z, \tau_{peak})$ as a function of distance along a $L = 10$ cm long structure (taken to be the measured 3.5 ns delay index profile from Fig. 2(a)) for varying $w_0$ for peak intensity $I_0 = 3 \times 10^{17}$ W/cm$^2$. Here $\tau = t - z/v_g$ is a time coordinate local to the pulse, $v_g$ is the pulse group velocity, and $\tau_{peak}$ is the location of the pulse peak. Figure 4(b) plots the energy transmission efficiency $T(z) = \varepsilon_i^{-1} \int d\tau d^2\mathbf{r}_\perp I(\mathbf{r}_\perp, z, \tau)$ vs. $z$, where $\varepsilon_i = 0.53\, I_0 \pi w_0^2 \tau$ is the pulse energy at the entrance to the index structure ($z = 0$) and the integral is over the pulse envelope and the index structure cross section. We will later discuss the physical origin of the oscillations and transmission decay in Figs. 4(a) and (b).

The structure cross section at $z = 2$ cm ($8z_0$) is shown in Fig 4(c) before and after injection of pulses with the varying waists from Fig. 4(a) and (b); it is clear that the plasma waveguide profile ultimately generated is largely dictated by the initial neutral hydrogen profile (light green points) and *not* the injected beam waist. This is consistent with the measurements of Fig. 3(a). The computed [14,21] lowest order quasi-bound mode for Fig. 4(c)'s similar plasma waveguide profiles is shown overlaid on the plot; it is near-Gaussian with a $1/e^2$ intensity waist $w_{ch} \sim 20$μm, matching the $w_0 \sim 20$μm plot of Fig. 4(a), which has the least intensity oscillation. This suggests that injected pulses that are not matched to the ultimate waveguide profile could couple into the profile's higher order modes. Indeed, the source of the strong oscillation for the $w_0 \sim 15$ μm and $w_0 \sim 25$ μm cases in Fig. 4(a) is beating between the first two quasi-bound modes excited when the injected pulse $w_0$ is not matched to the $w_{ch}$ of the ultimate self-generated guide profile (Fig. 4(c)). These modes correspond to $(p, m) = (0, 0)$ and $(1, 0)$, where $p$ and $m$ are radial and azimuthal mode indices, and where $m = 0$ owing to negligible azimuthal asymmetry in the $J_0$ pulse-excited index structure. The intensity profiles of the $(0, 0)$ and $(1, 0)$ modes are shown in Fig. 4(e) (top two sub-panels). The mode intensity ratio is $\mathcal{R} = |a_{0,0}/a_{1,0}|^2 \sim 43$, where $a_{0,0}$ and $a_{1,0}$ are the modal expansion coefficients for the waveguide profiles of Fig. 4(c), and their computed wavenumber difference is $\Delta\beta = \beta_{0,0} - \beta_{1,0} = 2.3$ cm$^{-1}$ [14,21]. Beating between these modes would have a period $\Lambda = 1/\Delta\beta = 4.3$ mm and a linear modulation visibility $v = (I_{max} - I_{min})/(I_{max} - I_{min}) = 2\sqrt{\mathcal{R}}/(\mathcal{R} + 1) = 0.30$. The modulation period corresponds very closely to the oscillation period of the 'non-matched' curves of Fig. 4(a), while the modulation visibility is qualitatively consistent with what is really a transient nonlinear process involving ionization and plasma lensing, along with mode beating. The bottom two sub-panels of Fig. 4(e) show the intensity profiles at the maximum and minimum of the oscillations in Fig. 4(a) (red and green dots).

It is interesting to note in Fig. 4(a) that for the unmatched injection pulse with $w_0 = 25$ μm, the peak intensity oscillations grow to as much as ~$2.7\ I_0$ before they settle down to a steady level of ~$1.5\ I_0$, followed by a smooth erosion decay to the channel exit. By contrast, the $w_0 = 15$ μm pulse oscillations never exceed $I_0$ and settle down to a steady level of ~$0.5\ I_0$ before decaying smoothly to the exit. This is from injected energy being channeled into a spot $w_{ch}$, transversely compressing or expanding the pulse. This predicts an intensity enhancement $(w_0/w_{ch})^2 \sim (25/20)^2 \sim 1.6 \times$ for the $w_0 = 25$ μm pulse injection, and an intensity reduction $(w_0/w_{ch})^2 \sim (15/20)^2 \sim 0.6 \times$ for the $w_0 = 15$ μm pulse injection. These estimates agree well with the simulations.

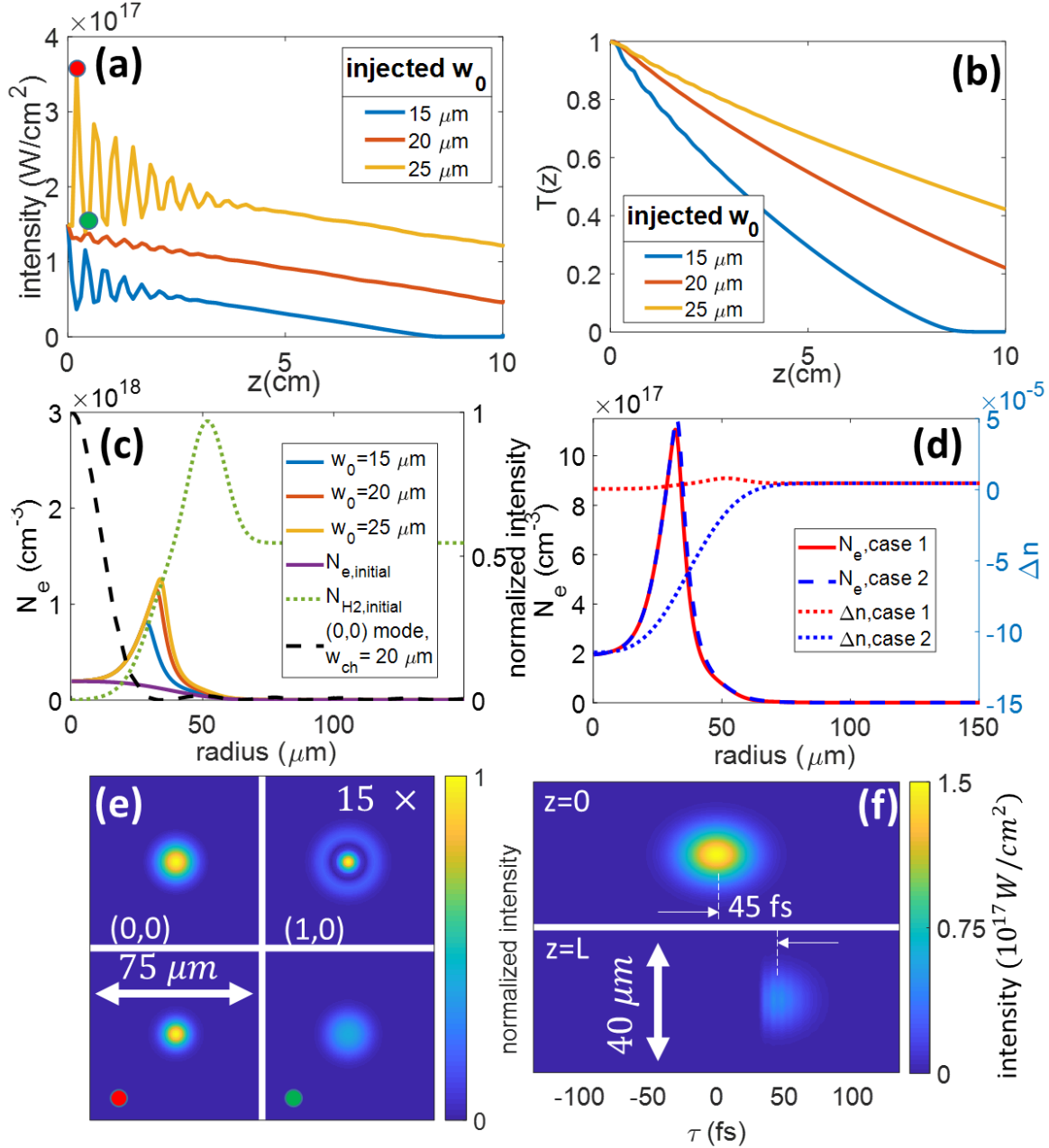

**Figure 5.** Peak input intensity at $z = 0$ is $I_0 = 1.5 \times 10^{17}$ W/cm$^2$. The panel descriptions are the same as in Fig. 4, except that in (e), the (1,0) mode is shown scaled in intensity by $15 \times$.

As discussed earlier, the on-the-fly generation of the plasma waveguide cladding is driven by the portion of the pulse's leading edge in excess of the hydrogen ionization threshold. Thus, earlier portions of the pulse below the ionization threshold will diffract away, and later portions, short of achieving full cladding confinement, will leak away until the cladding height and width is sufficient to stem further losses. Pulse head erosion is seen in Fig. 4(f), which compares (for the matched input $w_0 = 20$ μm) the input pulse at $z = 0$ and the pulse exiting the structure at $z = 10$ cm. In addition, the exit pulse centroid delay (~40 fs) is consistent with the group velocity delay $\Delta \tau = L(v_g^{-1} - c^{-1}) \sim 45$ fs, where the group velocity of the pulse in the plasma waveguide [21] is $v_g = (\partial \beta_{0,0}/\partial \omega)^{-1} = c[1 - N_{e0}/2N_{cr} - 2/(kw_{ch})^2]$. Here, $N_{e0} = 2 \times 10^{17}$ cm$^{-3}$ and $w_{ch} = 20$ μm (from Fig. 4(b)), $k$ is the laser vacuum wavenumber, and $N_{cr} = 1.7 \times 10^{21}$ cm$^{-3}$ is the critical density for $\lambda = 800$ nm. We note that the self-generated waveguides shown thus far are tightly confining, with $L_{1/e} > 30$ cm for those simulated Fig. 4(c) and $L_{1/e} > 1.4$ m for the 88 mJ-generated guide of Fig. 3(a).

An important question arises: is there a class of initially prepared refractive index profiles that is best-suited to support self-waveguiding? Thus far in Fig. 4, we presented simulations corresponding to our experimental index profiles: a low density plasma core surrounded by an enhanced density neutral annulus. In Fig. 4(d) we simulate self-waveguiding in two other types of index profiles. In case 1, the initial refractive index profile $\Delta n_{case1}(r)$ corresponds to a depressed on axis neutral density surrounded by an annulus of enhanced density corresponding to the profile shown in Fig. 4(c). There is no plasma contribution. The other case, $\Delta n_{case2}(r)$, removes the enhanced neutral density annulus and keeps Fig. 4(c)'s central plasma core and neutral density depression. We note that the central concavity of both $\Delta n_{case1}$ and $\Delta n_{case2}$ profiles would ensure defocusing of a low intensity injected pulse, even more strongly for case 2. However, end-injection of the $I_0 = 3 \times 10^{17}$ W/cm$^2$, $w_0 = 20$ μm pulse into either the $\Delta n_{case1}(r)$ or $\Delta n_{case2}(r)$ structure yields the electron density profiles at $z = 2$ cm overlaid on the plot; they are nearly identical. Full propagation runs (not plotted here) show indistinguishable self-waveguiding—ionization occurs early enough that the remainder of each pulse sees the same plasma waveguide. We conclude that the most important feature of a prepared index profile is an adequately deep neutral density gas depression, where for hydrogen gas, the depth must exceed $\Delta N_{H2} > (2\pi r_e w_{ch}^2)^{-1}$ to confine a mode $w_{ch}$ [21], where $r_e$ is the classical electron radius. The presence of a neutral shock wall (as included in case 1) is unnecessary in general; however, under some circumstances it will reduce propagation leakage by providing higher and wider cladding. Plasma is required only to the extent that it contains the laser-provided thermal energy that drives the initial cylindrical blast expansion.

Figure 5 repeats the simulations of Fig. 4 except at a lower peak injected intensity, $I_0 = 1.5 \times 10^{17}$ W/cm$^2$. Here, the deleterious effects of a 2× intensity reduction are evident. The basic propagation features remain, such as the mode beating for unmatched injected beams, but we also observe the following: $(i)$ The peak intensity $I(r = 0, z, \tau_{peak})$ and transmission $T(z)$ decline much faster with $z$ (Fig. 5(a) and 5(b)). $(ii)$ While the electron profiles in 5(c) still follow the neutral density profile (light green points) the peak and width of the self-generated electron density cladding are smaller, especially for the $w_0 = 15$ μm injected pulse. This accounts for the much faster leakage rate seen in Fig. 5(b)—in fact, the $w_0 = 15$ μm pulse does not even survive to $z = 10$ cm. $(iii)$ The intensity ratio of the (0,0) mode to the (1,0) mode is much smaller, $\mathcal{R} = |a_{0,0}/a_{1,0}|^2 \sim 15$ (for the $w_0 = 20$ μm –induced electron density profile in Fig. 5(c)), giving larger linear modulation visibility $v = 0.48$. For all 3 profiles $\Delta \beta = \beta_{0,0} - \beta_{1,0} = 2.3$ cm$^{-1}$ as before, giving rise to the same modulation periods seen in Figs. 4(a) and 5(a). The pulse energy redistribution effect is likewise seen in Fig. 5(a), with a peak intensity enhancement (after oscillations settle down) of $\sim 1.3 \times$ (for $w_0 = 25\ \mu m$) and a $\sim 0.5\times$ reduction (for $w_0 = 15\ \mu m$), reduced effects owing to the weaker confinement of the self-generated guides. $(iv)$ Pulse head erosion, as seen in Fig. 5(f), is more significant, as a larger portion of the leading edge is below the self-trapping threshold. This has the effect of moving the centroid of the exit pulse farther back.

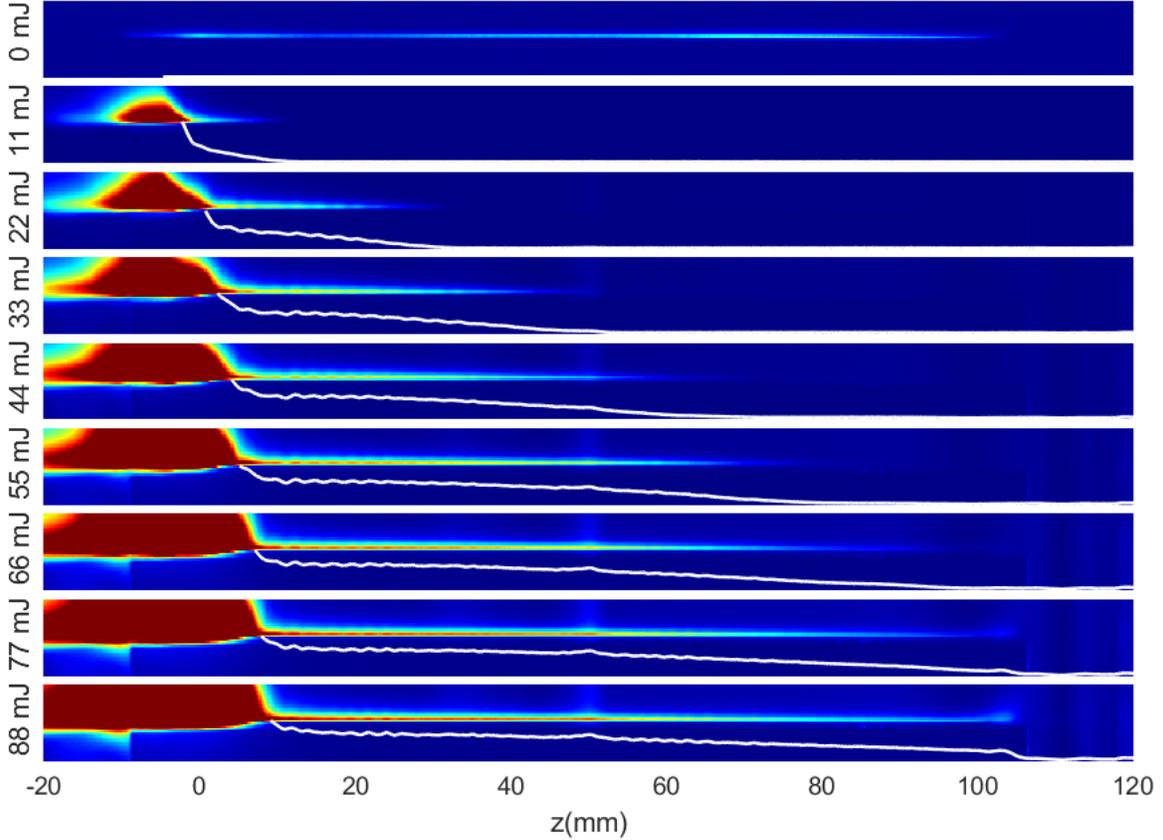

**Figure 6.** Full length hydrogen plasma fluorescence (recombination) images of self-waveguiding of $\lambda = 800$ nm, 40 fs pulses of increasing energy in a 10 cm long index structure generated in an elongated supersonic hydrogen jet by a 100 mJ $J_0$ pulse. $\Delta t_{injection} = 2.5$ ns into the profiles of Fig. 3(a). The pulse propagates left to right. Below each image is a central lineout (white curves). The decreasing fluorescence with z indicates pulse head erosion of the self-waveguiding pulse. The bump at $z \sim 50$ mm is from light scattering from an optical support post in the experimental chamber. The saturated (brown) image at the jet entrance is from the injected laser pulse overfilling the initial index structure. The slight bump at the channel exit is from ionization of neutrals by the exiting pulse. The top panel shows fluorescence induced by the $J_0$ pulse alone.

We now show results from our guiding experiments in the 10 cm long hydrogen jet. Figure 6 shows a sequence of hydrogen plasma recombination (fluorescence) images of the full jet for increasing injected pulse energy, for $\Delta t_{inj} = 2.5$ ns (into the 2.5 ns delay index profile of Fig. 2(a), resulting in the plasma profiles of Fig. 3(a)). The $J_0$ line focus is 0.8mm above the nozzle's slot orifice, where the local atomic hydrogen density is $N_0 = 3.2 \times 10^{18}$ cm$^{-3}$. The z-dependent fluorescence is proportional to the local radial integral of the electron density, $F(z) \propto \int_0^{r_{max}} dr 2\pi r N_e(r, z)$, and tracks the z-dependent free electron charge generated by the guided pulse; it is a qualitative measure of the axial extent of guiding.

As the injected laser energy increases, the pulse propagates progressively longer distances because a smaller fraction of its leading edge is below the thresholds for hydrogen ionization and self-trapping. That fraction continually erodes from the front of the pulse, reducing its energy until a self-trapping structure can no longer be generated and the remaining pulse diffracts away. We see that for injected energy >75 mJ, the pulse makes it to the end of the 10 cm long structure and then expands, where it excites a small burst of fluorescence in residual neutral gas at the exit. The much larger fluorescence signal at the entrance (shown saturated) may be from ionization of the index structure by the wings of the injected beam mode. The oscillations seen in the simulations of Figs. 4(a) and 5(a) may contribute to these wider wings, as seen in Figs. 4(e) and 5(e). Below each fluorescence image is a central lineout (white curves), from which an erosion rate of ~6 mJ/cm is estimated from the downward slope. This is mainly a diffraction and leakage loss: absorption is negligible, as the laser energy expended on ionization is < ~1 mJ for all guides measured in our experiments. Measurable end mode transmission occurred only when the fluorescence showed 'burn through' to the end of the jet. The improved transmission with increasing pulse energy is also seen in the simulations by comparing $T(z)$ in Figs. 4(b) and 5(b). These curves show an energy erosion rate between 7.5 mJ/cm (Fig. 4, $I_0 = 3 \times 10^{17}$ W/cm$^2$, $w_0 = 25$ $\mu m$) and 4 mJ/cm (Fig. 5, $I_0 = 1.5 \times 10^{17}$ W/cm$^2$, $w_0 = 25$ $\mu m$), consistent with our measured erosion rate of ~6 mJ/cm.

A feature common to the fluorescence images for injected pulse energy > 25 mJ is a prominent axial modulation. In Fig. 7, we plot central lineouts from fluorescence plots for injected 88 mJ pulses at a range of delays $\Delta t_{inj}$, with their z-averaged modulation periods Λ indicated on the plot. Each of these delays corresponds to an index profile plotted in Fig. 3(d). Based on the simulations and analysis of Figs. 4 and 5, we attribute the modulations to beating between the (0,0) and (1,0) modes. Using the profiles of Fig. 3(d), we calculate the quasi-bound mode wavenumber spectrum [14, 21]and determine $Λ_{sim} = (\Delta\beta)^{-1}$, where $\Delta\beta = \beta_{0,0} - \beta_{1,0}$. These values are added to the panels in Fig. 7 and show mostly good agreement with the measured z-averaged periods Λ (except for the $\Delta t_{inj} = 5$ ns case, for reasons unclear).

For successful self-waveguiding of intense laser pulses, the self-generated guide must be able to support at least one low loss quasi-bound mode, as in Figs. 4(c) and 5(c). An approximate condition on the waveguide profile for trapping and guiding of a $(p, m)$ mode is

$$\Delta N_e > (2p + m + 1)^2/\pi r_e w_{ch}^2, \qquad (1)$$

where $\Delta N_e \approx N_e(r = w_{ch}) - N_e(r = 0)$ and $w_{ch}$ is the radius of the corresponding fundamental mode [21]. The requirement of minimum $\Delta N_e$ implies a threshold initial hydrogen gas density for guiding. This is seen in Fig. 8(a) (for injection pulse energy 88 mJ and $\Delta t_{inj} = 2.5$ ns), which shows a sharp onset of guiding over the full 10 cm channel for $N_{H2} > 10^{18}$ cm$^{-3}$. Shown are plots of pulse energy exiting the guide and transmission efficiency (pulse energy at guide exit /energy at entrance), along with exit modes measured for low, intermediate, and higher hydrogen densities. At the onset of guiding, only the (0,0) mode exits the guide. As H$_2$ density increases, the contribution of the (1,0) mode becomes more significant, consistent with Eq. (1), which implies that density should increase by ~4 × for $p = 1$ mode confinement. The presence of the (1,0) mode is consistent with the mode beating seen in Fig. 6. Injected pulse transmission vs. energy is shown in Fig. 8(b) (for $\Delta t_{inj} = 2.5$ ns and $N_{H2} = 3 \times 10^{18}$ cm$^{-3}$), where the low end of the energy axis is the minimum needed for burn-through of the 10 cm long structure. The increased transmission at high energy (approaching ~10%) under these conditions reflects the reduced fractional erosion of the head of the pulse. The highest transmitted energy of 23 mJ corresponds to a peak self-waveguided exit intensity of $1.5 \times 10^{17}$ W/cm$^2$, in the weakly relativistic regime (normalized vector potential $a_0 \sim 0.3$). Upstream from the exit were even higher intensities, as is clear from Figs. 4(a,b) and 5(a,b).

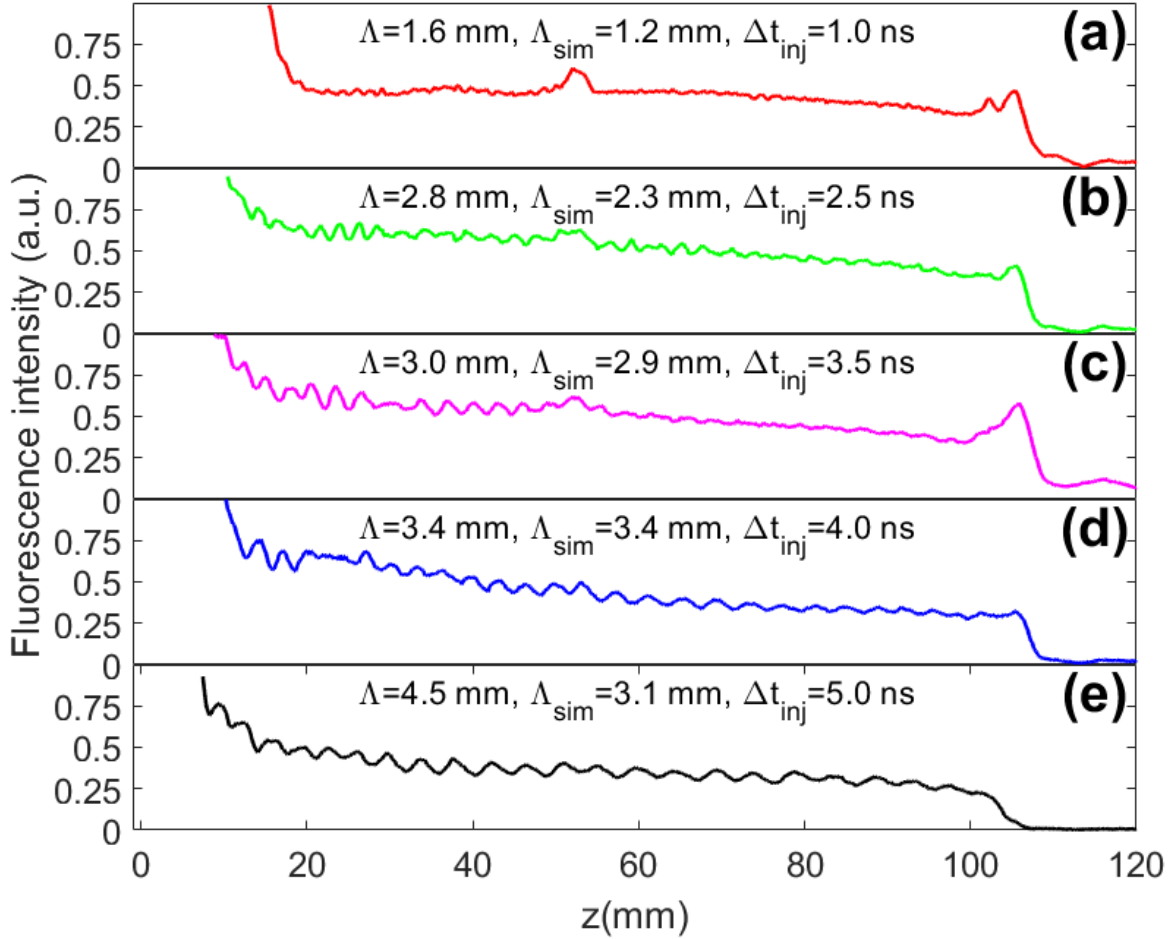

**Figure 7.** Full length hydrogen plasma fluorescence (recombination) image lineouts of self-waveguiding (of $\lambda = 800$ nm, $w_0 = 25$ μm, 40 fs, 88 mJ pulses) in a 10 cm long index structure generated in an elongated supersonic hydrogen jet by a 100 mJ $J_0$ pulse. Here the injection delay $\Delta t_{inj}$ is varied from 1 ns to 5 ns, slightly changing the self-generated plasma waveguide (as seen in Fig. 3(d)) and its quasi-bound modes and wavenumbers $\beta_{0,0}$ and $\beta_{1,0}$, leading to a variation in the mode beating period $\Lambda_{sim} = (\Delta\beta)^{-1}$, where $\Delta\beta = \beta_{0,0} - \beta_{1,0}$. The simulated period $\Lambda_{sim}$ and measured profile-average period $\Lambda$ are in reasonable agreement. The bump at z~50 mm is from light scattering from an optical support post in the experimental chamber.

The simulations in Figs. 4 and 5 show a strong sensitivity of $T(z = L)$ to $I_0$ and $w_0$, which is likely reflected in our transmission measurements. Given the uncorrected astigmatism from the 0.5 m spherical focusing mirror, we attribute the mismatch between the transmission measurements and the simulations of Figs. 4(b) and 5(b) to the uncorrected mode. Nevertheless, our results are reasonably consistent with the simulations. Importantly, it is clear that higher intensity injected pulses will experience much higher transmission fractions.

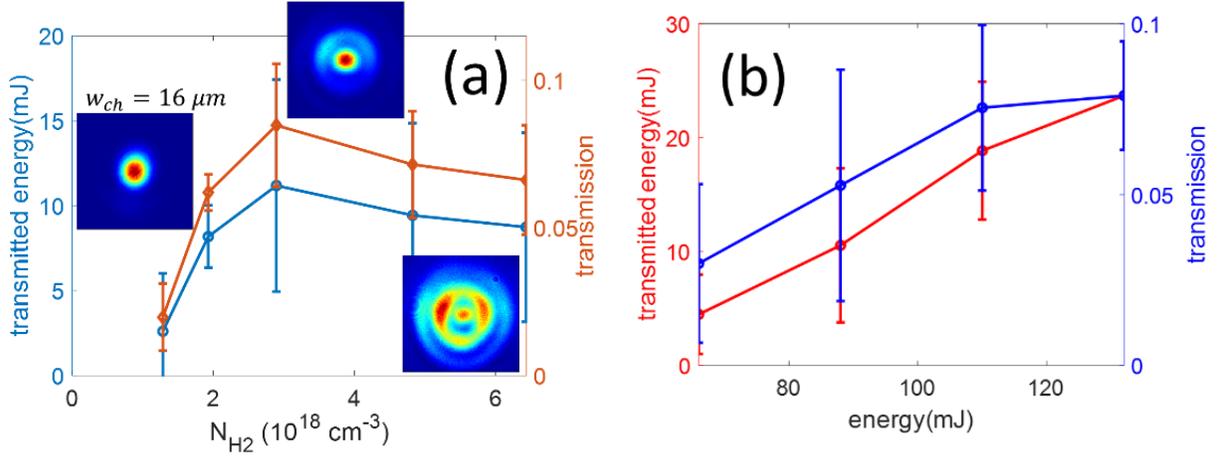

**Figure 8. (a)** Plot showing hydrogen density threshold for self-waveguiding (injection pulse energy 88 mJ, $\Delta t_{inj} = 2.5$ ns). As density increases, both (0,0) and (1,0) modes are seen at the channel exit, with the contribution of the (1,0) mode increasing. The slightly reduced transmission at higher density may be from leakage of the (1,0) mode. **(b)** Injected pulse transmission vs. energy (for $\Delta t_{inj} = 2.5$ ns and $N_{H2} = 3 \times 10^{18}$ cm$^{-3}$), where the low end of the energy axis is the minimum needed for burn-through of the 10 cm long structure. Peak self-waveguided intensity exiting the guide is $1.5 \times 10^{17}$ W/cm$^2$.

The relatively low transmission efficiency measured here for pulses in the $\sim 10^{17}$ W/cm$^2$ range should not be considered as an argument against implementing the self-waveguiding method. Rather, it could be considered the cost of generating a plasma waveguide for a follow-on pulse of much higher intensity. Such a follow-on pulse can be used, for example, for LWFA. To illustrate this operating scenario, we performed a 3-pulse experiment in our 10 cm gas jet: a $J_0$ pulse ($\lambda = 800$nm) generated the initial refractive index structure, a pump pulse ($\lambda = 800$nm) injected into the structure at $\Delta t_{inj} = 2.5$ ns generated the plasma waveguide, and a probe pulse ($\lambda = 400$nm) followed the waveguide generation pulse at delay ~5 ps. The probe pulse mode was not optimized and overfilled the waveguide entrance. Figure 9 is a plot of probe pulse transmission as a function of pump pulse energy. Above the curve are shown $\lambda$=400 nm probe output mode images and below the curve are the corresponding self-waveguiding exit mode images at several injected pump pulse energies. Most notable is that all modes, for both pump and probe at $\lambda = 800$nm and $\lambda = 400$nm, are similar, with $w_{ch} = 16 - 18$ μm. This is strong evidence of two important features: First, it clearly reemphasizes that the prime determiner of the plasma waveguide structure left by a self-waveguiding pulse is the initial neutral density profile induced by the index-structuring $J_0$ pulse. And second, it verifies that the mode structure of a given plasma waveguide is wavelength-independent [21].

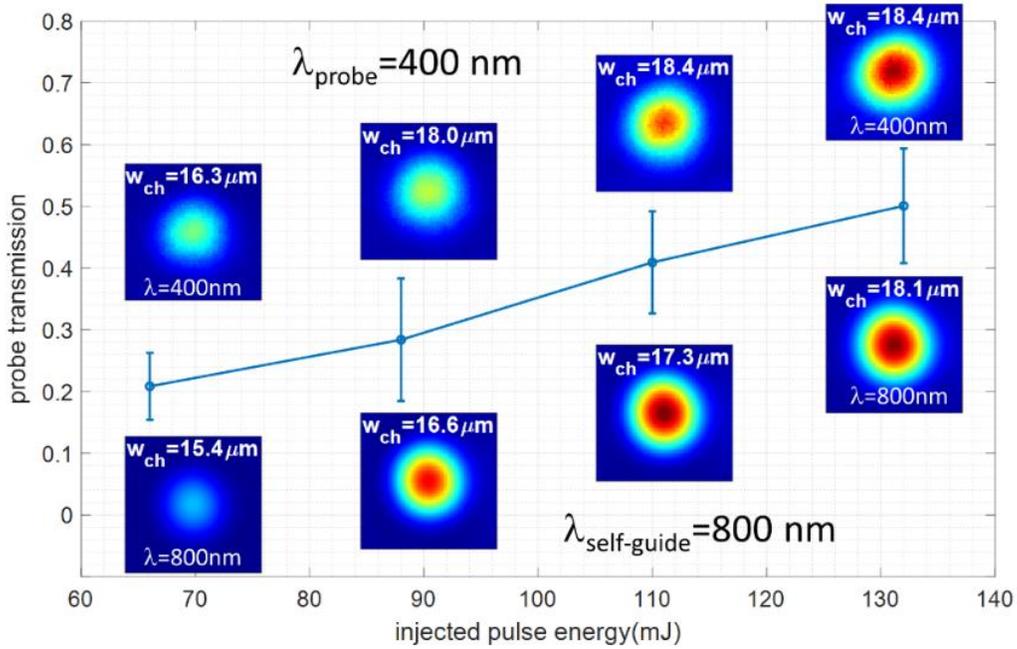

**Figure 9.** Pump-probe experiment, where the pump is the self-waveguiding pulse (λ=800 nm) injected at $\Delta t_{inj} = 2.5$ ns after the 100 mJ $J_0$ pulse, followed at 5 ps delay by a probe pulse (λ=400 nm). The corresponding initial index structure is plotted in Fig. 2(a), and its generated waveguide is shown in Fig. 3(a). The self-waveguided λ=800 nm exit modes are shown below the curve, with the corresponding λ=400 nm probe exit modes shown above the curve. The mode invariance ($w_{ch} = 16 - 18$ μm) illustrates that (i) the resulting plasma waveguide structure is nearly independent of the energy of the self-waveguiding pulse and (ii) the plasma waveguide mode structure is wavelength-independent.

## 4. CONCLUSIONS

We have presented results from detailed experiments and simulations showing that an intense ultrashort laser pulse can generate its own plasma waveguide as it propagates through a long refractive index structure imprinted in neutral gas. The necessary feature of such a structure is a neutral gas density minimum on axis. We call this process 'self-waveguiding', which is distinct from the well-known process of relativistic self-focusing followed by self-guiding, in which laser ponderomotive force electron cavitation. Self-waveguiding requires much lower intensities than for relativistic self-guiding, with the main requirement that the transverse wings of the pulse exceed the ionization threshold of the gas. For hydrogen, this means that for guided modes in the range studied in this paper, $w_{ch} \sim 20$ μm, the peak injected intensity should be at least $\sim 10^{17}$ W/cm² for intensity in the wings to be $> \sim 10^{14}$ W/cm² to ionize neutral hydrogen.

For conditions explored in our experiments, the initial refractive index structure was generated using a $J_0$ Bessel beam focused in a 10 cm long hydrogen gas jet. This pulse generated a 10 cm long, thin optical field ionized plasma that launched a cylindrical shock wave into the peripheral neutral hydrogen, forming the index structure into which weakly relativistic pump pulses were end-injected. We demonstrated self-waveguiding over 10 cm, equivalent to $\sim 100$ Rayleigh ranges at λ=800 nm and a spot size $w_{ch} = 16$ μm, with peak intensity $1.5 \times 10^{17}$ W/cm² exiting the waveguide. Importantly, we find that the initiating plasma is important only to the extent that it provides the thermal energy to drive the on-axis density depression. Any other method that could drive such a depression would work. Provided sufficient laser energy is available, our self-waveguiding results are easily extendable to much longer interaction lengths, and therefore waveguide generation lengths, using a longer gas jet or employing a long gas cell. The waveguides of this experiment were generated at a cost of approximately ~1.3 J/m ( ~7 mJ/cm for the $J_0$ pulse and ~ 6 mJ/cm erosion cost for the self-waveguiding pulse), a modest energy expenditure for many current high energy short pulse laser systems .

While our measured transmission of self-waveguiding pulses is low, transmission quickly increases with laser energy. In any case, the process is very well-suited for generating waveguides for follow-on pulses of much higher intensity for applications such as LWFA. We demonstrate this principle in a pump-probe experiment, where the pump is a λ=800nm self-waveguiding pulse and the probe is a low energy λ=400nm pulse. On the other hand, our results indicate that higher intensity self-waveguiding pulses approaching $10^{18}$ W/cm$^2$ could generate the waveguide, experience high transmission, and drive LWFA in the quasilinear regime  [4].

Finally, as a means for plasma waveguide generation, we compare the scheme of this paper to another approach of our group. In recent work [11], we have shown that very long (30 cm), low density plasma waveguides can be formed using two Bessel beam pulses: the first pulse, a $J_0$ beam, generates the waveguide core, and the second pulse, $J_8$ or $J_{16}$, generates the plasma cladding using its main and subsidiary rings.  To generate the plasma waveguide profile of Fig. 3(a) (for which $w_{ch} = 20$ μm) over 10 cm using the two-Bessel method, we estimate [14]an energy requirement of $\sim 5 - 10$ mJ/cm, in the same range as the ~6 mJ/cm estimated for the self-waveguiding pulses of the current paper. As the required $w_{ch}$ increases further, the energy required in the self-waveguiding pulse must scale as $\sim w_{ch}^2$ to maintain sufficient intensity in the wings to ionize the neutral gas. Maintaining that intensity in the main ring of the higher order Bessel beam pulse requires sublinear energy scaling $\sim w_{ch}^{0.63}$ [14].

## ACKNOWLEDGEMENTS

The authors thank M. Tomlinson for experimental assistance, J. Griff-McMahon for help with the simulations, and A. Picksley, A. Ross, and S. Hooker for ongoing discussions. This work was supported by the US Department of Energy (DESC0015516); National Science Foundation (PHY1619582); Air Force Office of Scientific Research (FA9550-16-1-0121, FA9550-16-1-0284); Office of Naval Research (N00014-17-1-2705, N00014-20-1-2233).

# Supplementary material for "Self-waveguiding of relativistic laser pulses in neutral gas channels"

## 1. Calculation of quasi-bound modes

As discussed in detail in [1], plasma waveguides with a finite thickness cladding have modes that are quasi-bound or leaky. A sufficiently well-bound leaky mode has an exponential decay length that is long compared to other scale lengths of interest, such as the dephasing length or pulse decay due to energy depletion in laser wakefield acceleration.

Here, quasi-bound or leaky modes were found for azimuthally averaged density profiles by solving the Helmholtz equation for the field $E(r,z) = \mathcal{E}(r)e^{i\beta z}$ in cylindrical coordinates,

$$d^2\mathcal{E}/ds^2 + s^{-1}d\mathcal{E}/ds + (n^2 - \beta^2/k_0^2)\mathcal{E} = 0 \ , \tag{S.1}$$

where $k_0$ is the vacuum wave number, $\beta$ is the waveguide propagation wavenumber, $s = k_0 r$, and $n(r)$ is the refractive index profile derived from the azimuthally averaged plasma profile. To map out the quasi-bound modes for a given index profile $n(r)$, we repeatedly solve Eq. (S.1), scanning $\beta' = \beta/k_0$ and plotting $\eta(\beta') = (|\mathcal{E}_{vacuum}|^2 A)^{-1} \int_A |\mathcal{E}|^2 dA$, where the integral is over the waveguide cross section. The main peaks in $\eta(\beta')$ identify the quasi-bound modes, with the full width at half maximum $\Delta\beta$ of these peaks giving the axial ($z$) $1/e$ attenuation length of intensity $L_{1/e} = \Delta\beta^{-1}$ [1].

## 2. Pulse propagation simulations

The simulations of beam propagation in this document were performed using a 2D UPPE[2] implementation called YAPPE ('Yet Another Pulse Propagation Effort'). UPPE ('Unidirectional Pulse Propagation Equation') is a system of ordinary differential equations (ODEs) of the form

$$\frac{\partial}{\partial z} A_{k_\perp}(\omega, z) = iQ_{k_\perp}(\omega) 2\pi P_{k_\perp}(\omega, z) e^{-i\left(k_z - \frac{\omega}{v_g(\omega)}\right)z} \ . \tag{S.2}$$

In Eq. (S.2), $A = A_{k_\perp}(\omega, z)$ is an auxiliary field related to the Fourier transform of the optical field by $E = Ae^{ik_z z}$. The spectrum of radial spatial frequencies $k_\perp$ indexes a system of ordinary differential equations, which is solved using a GPU implementation of MATLAB's ODE45 function. $P_{k_\perp}(\omega, z)$ is the nonlinear polarization of the medium including (for our simulations in this paper): third-order optical nonlinearities in $H_2$, ionization losses, and the optical plasma response, $\omega$ is angular frequency, $v_g(\omega)$ is the group velocity of the pulse as a function of frequency, $k_z = ((\omega/v_g(\omega))^2 - k_\perp^2)^{1/2}$ is the longitudinal (z-direction) spatial frequency, and $Q_{k_\perp}(\omega) = \omega/ck_z$. The ionization rate is computed using an ADK module with neutral density saturation[3,4]. The perturbed index of refraction due to the $H_2$ neutral density gradient is also included as a part of the nonlinear polarization. The strength of the third-order nonlinearity (proportional to $n_2$) of hydrogen[5] is scaled with the spatiotemporally variable $H_2$ neutral density such that it goes to zero when all neutrals are ionized.

# 3. Interferometric measurement of plasma density profiles

(a) *1D extraction for azimuthally symmetric profiles.* Transverse profiles of electron density and neutral density were extracted from 2-color interferograms generated by ultrashort $\lambda = 400\ (800)$ nm, $70\ (40)$ fs probe pulses, imaged from the plasma through a femtosecond shearing interferometer and onto a CCD camera. Each profile was constructed using 200 shots (frames). The analysis steps are as follows. First, the 2D interferometric phase shift $\Delta\Phi(y,z)$ was extracted from each frame using standard techniques[1], where $z$ and $y$ are coordinates along and transverse to the optical axis. In each column (fixed $z$), background subtraction was performed by fitting the phase in the perturbation-free part of the frame to a polynomial and interpolating over the full frame. An average phase shift $\overline{\Delta\Phi(y)}$ was then computed over 500 columns of the extracted phase and then over 200 shots, where the phase shift profile is very stable from shot to shot. This procedure resulted in phase shift resolution of ~4 mrad (assuming a minimum signal-to-noise ratio of 20 dB and measured standard deviation background noise of 0.3-0.4 mrad). $\overline{\Delta\Phi(y)}$ was then Abel inverted to recover the respective refractive index shift profiles density.

(b) *Two-color density profile extraction and error analysis.* The neutral gas and electron density profiles in Fig. 2(a) are measured using two-color interferometry with $\lambda = 400$ nm and $\lambda = 800$ nm probe beams. The probe phase shift for colour $i$ is $\Delta\phi_i = (2\pi/\lambda_i)\int dx \left(\frac{\Delta N_g}{N_{STP}}\delta n_{g,STP}^{(i)} - \frac{N_e}{2N_{cr}^{(i)}}\right) = \int dx\ (\delta\phi_g^{(i)} + \delta\phi_e^{(i)})$ for $\lambda_1 = 400$ nm and $\lambda_2 = 800$ nm. $\Delta N_g$ is the change in hydrogen molecular density profile, $(1 + \delta n_{g,STP}^{(i)})$ is the refractive index of hydrogen at STP for $\lambda = 400$ nm and $\lambda = 800$ nm [6], $N_e$ is the electron density profile, and $N_{cr}^{(i)}$ is the critical density at $\lambda_i$. Separate phase shift contributions of neutral hydrogen and plasma are then extracted and Abel inverted by a matrix method to extract the neutral gas and plasma density profiles [7].

The uncertainty in the Abel inverted plots is determined by error propagation in the Abel inversion process [8]. If the Abel transform is written as a matrix operation, then the measurement is expressed as $G = AF$, where $F$ and $G$ are the phase shift density and the measured phase shift respectively, and $A$ is the Abel transform matrix. If we write the covariance matrices of $F$ and $G$ as $var(F)$ and $var(G)$, then it can be shown [8] that $var(F) = A^{-1}\ var(G)(A^{-1})^T$. Usually, the elements of $G$ are independent so $var(G)$ is diagonal, and it is reasonable to write $var(G) = \sigma^2 I$, where $\sigma$ is the phase noise in the interferograms. Therefore, $var(F) = \sigma^2(A^T A)^{-1}$. To calculate $\sigma^2$, we use variance of the deviation between the measured phase shift and reconstructed phase shift $\phi_{exp} - \phi_{rec}$ as a conservative estimate of the uncertainty.

(c) *Quasi-2D extraction for azimuthally asymmetric plasma profiles.* For the cylindrically symmetric plasma generated by $J_0$ pulses, the procedure in 3(a) was sufficient. However, the plasma formed by the self-waveguiding beam is frequently asymmetric due to pointing and position offset between the two beams in the experiment. Standard Abel inversion often results in unphysical plasma density near the axis due to asymmetric phase shift.

We extract plasma density in two steps. First, the plasma column formed by the $J_0$ beam before the self-waveguiding beam is extracted with standard Abel inversion. The corresponding phase shift $\phi_0$ is subtracted from the measured phase shift $\phi_{tot}$, giving $\phi_w = \phi_{tot} - \phi_0$ as the contribution of the self-waveguiding beam. Assuming the density profile can be approximated as $F(r,\phi) = f(r)\sum_{m=0}^{3} c_m \cos(m\phi + \Delta\phi_m)$, $\phi_w$ is extracted as $\mathrm{argmin}_F \int |\mathcal{A}[F] - \Delta\Phi(y)|^2 dy$, subjected to $F(r,\phi) \geq 0$. The radial profile $f(r)$ was fit by a Gaussian basis set to improve smoothness and account for the limited resolution in interferometry, as detailed in[9]. We have no information on $\Delta\phi_m$, so we assumed $\Delta\phi_{1,3} = \pi/2$ and $\Delta\phi_2 = 0$ for convenience, as the choice of $\Delta\phi_m$ doesn't affect the azimuthally averaged density profile $f(r)$. The optimization used the built-in MATLAB function *fmincon*, and $f(r)$ was initialized with random noise. The criterion of convergence was set as the normalized error $\epsilon = \int |\mathcal{A}[F] - \Phi_w(y)|^2 dy / \int |\Phi_w(y)|^2 dy < 0.001$. The optimization was initialized with random noise as $f(r)$ and ended with the same $f(r)$ for each case presented in the main text. Therefore, we conclude that uniqueness was achieved. The plasma density profile extracted from $\phi_0$ and $\phi_w$ were combined as shown in figure 3(a)(b)(d). In figure 3(c), since information on $\phi_0$ is not available, we assume $\phi_0 = 0$ and repeat the rest of the procedure.

**4. Plasma waveguide generation: comparison of energy requirements for self-waveguiding vs. the two-Bessel pulse method**

The cost per unit length for the two-Bessel pulse method [11] in producing a plasma profile similar to the one shown in Fig. 3(a) of the main paper was computed using a $J_q$ Bessel beam with $\alpha = 3°$, $\tau_{FWHM} = 50\,fs$ and a super Gaussian profile. We chose a 10 mm FWHM beam diameter and the order $q = 18$ to match the length and width of the self-waveguiding-generated plasma. As was shown in [11], the Bessel beam profile is unaffected by any preexisting plasma and the atomic density was assumed to be equal peak electron density in the self-waveguiding profile (blue curve in Fig. S1). The ionization rate was then calculated using a modified ADK model [10] and the z-averaged plasma profiles shown in Fig. S1 were found.

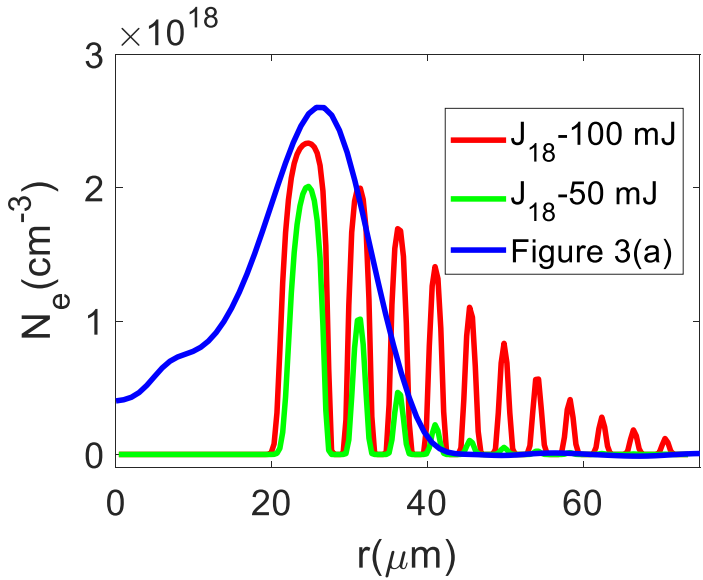

**Figure S1**. Energy requirement comparison. Blue curve: Plasma density profile shown in Fig. 3(a) of the main text. Red line: Plasma density profile generated by a $J_{18}$ Bessel beam pulse with 100 mJ energy, assuming a background hydrogen atom density $N_H = 2.5 \times 10^{18}\,\mathrm{cm}^{-3}$. Green line: Plasma density profile generated by a $J_{18}$ Bessel beam with 50 mJ energy with the same background $N_H$.

For a Bessel beam of order $q$, the intensity distribution can be written (for optical axis approach angle $\gamma = 2\alpha \ll 1$) as $I(r,z) = I(\rho_\gamma)(2\pi k\rho_\gamma \sin\gamma)J_q^2(kr\sin\gamma)$, where $\rho_\gamma = z\tan\gamma$ is radial position on the beam. If the plasma cladding is generated by the first ring of $J_q$, then we require $kr_{max}\sin\gamma = \xi_{q1}$, where $r_{max}$ is the peak location of the first ring, where $J_q'(\xi_{q1})=0$. So the z-dependent peak intensity along the first ring is $I(r_{max},z) = I(\rho_\gamma)(2\pi k\rho_\gamma \sin\gamma)J_q^2(\xi_{q1})$. Using the length of the focal line $L = R/\tan\gamma$ for a beam of radius $R$, the average ring intensity along the focal line is

$$\langle I(z)\rangle = \frac{1}{L}\int dz\, I(r_{max},z) = \frac{k\sin\gamma}{R}J_q^2(\xi_{q1})\int_0^R 2\pi\rho_\gamma I(\rho_\gamma)d\rho_\gamma$$

Using $\langle I(\rho)\rangle = (\pi R^2)^{-1}\int_0^R 2\pi\rho_\gamma I(\rho_\gamma)d\rho_\gamma$, we get $\langle I(z)\rangle = k\pi R\sin\gamma J_q^2(\xi_{q1})\langle I(\rho)\rangle$. Thus, to supply a focal line average threshold ring intensity $\langle I(z)\rangle_{th}$ to ionize hydrogen, the energy in the input beam (of pulse duration $\tau$) must be

$$\varepsilon \approx \langle I(\rho)\rangle\pi R^2\tau = \langle I(z)\rangle_{th}\frac{R\tau}{k\sin\gamma J_q^2(\xi_{q1})} \approx \langle I(z)\rangle_{th}\tau\frac{L}{kJ_q^2(\xi_{q1})}$$

The energy required per unit length is therefore

$$\frac{\varepsilon}{L} \approx \frac{\langle I(z)\rangle_{th}\tau}{kJ_q^2(\xi_{q1})}$$

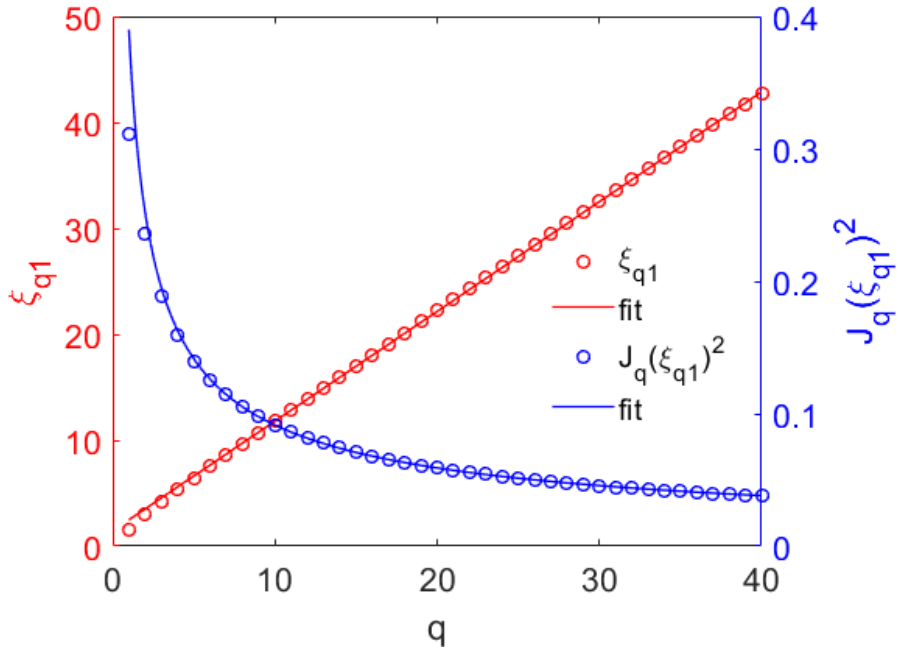

**Figure S2.** Computation of $\xi_{q1}$ and $J_q^2(\xi_{q1})$ vs. $q$ (red circles and blue circles). Fits are overlaid: $(\xi_{q1})_{fit} = 1.03q + 1.44$ and $(J_q^2(\xi_{q1}))_{fit} = 0.39q^{-0.63}$

Figure S2 plots $\xi_{q1}$ and $J_q^2(\xi_{q1})$ vs. $q$, with the fits $(\xi_{q1})_{fit} = 1.03q + 1.44$ and $(J_q^2(\xi_{q1}))_{fit} = 0.39q^{-0.63}$ overlaid. Taking $\xi_{q1} \approx q$ and using $w_{ch} = a\xi_{q1}$ where $a \lesssim 1$ gives $w_{ch} = qa/k\sin\gamma$ or $q = kw_{ch}\sin\gamma/a$. Therefore we can write $(J_q^2(\xi_{q1}))_{fit} = 0.39q^{-0.63} = 0.39(kw_{ch}\sin\gamma/a)^{-0.63}$, giving $J_q^{-2}(\xi_{q1}) \propto w_{ch}^{0.63}$. Therefore, the energy cost per unit length is

$$\frac{\varepsilon}{L} \propto w_{ch}^{0.63},$$

increasing sub-linearly with $w_{ch}$.